\documentclass[12pt,a4paper]{article}
\usepackage{epsf,epsfig,psfrag,latexsym,amssymb}
\newcommand{\be}{\begin{equation}}
\newcommand{\ee}{\end{equation}}
\newcommand{\bfi}{\begin{figure}}
\newcommand{\efi}{\end{figure}}
\newcommand{\bea}{\begin{eqnarray}}
\newcommand{\eea}{\end{eqnarray}}

\newcommand{\nc}{NCQED~}

\newcommand{\bec}{\begin{center}}
\newcommand{\ec}{\end{center}}

\def\d{\bf{\Delta}}

\def\amuhat{{\hat A_{\mu}}}
\def\anuhat{{\hat A_{\nu}}}

\def\xperp{\bf{x_{\perp}}}
\def\qperp{\bf{q_{\perp}}}
\def\ti{{\tilde I}}
\def\qoneperp{\bf{q_{1\perp}}}
\def\qtwoperp{\bf{q_{2\perp}}}
\def\qiperp{\bf{q_{i\perp}}}

\begin{document}
\begin{titlepage}
\title{Some Aspects of Scattering in (Non) Commutative Gauge Theories}
\author{}
\date{
Zafar Ahmed, Tapobrata Sarkar
\thanks{E--mail:~ shahi@ictp.trieste.it, tapo@ictp.trieste.it}
\vskip0.4cm
{\sl the Abdus Salam \\
International Center for Theoretical Physics,\\
Strada Costiera, 11 -- 34014 Trieste, Italy}}
\maketitle
\abstract{We study almost-forward scattering in the context of usual and
non-commutative QED. We study the semi-classical behaviour of particles undergoing
this scattering process in the two theories, and show that the shock wave picture,
effective in QED fails for NCQED. Further, we show that whereas in QED, there are no
leading logarithmic contributions to the amplitude upto sixth order, uncancelled
logarithms appear in NCQED.
}
\end{titlepage}
\pagenumbering{arabic}
\section{Introduction}

Scattering in gauge theories has been a topic of intense research ever since 
the formulation of Quantum Field Theory (QFT) several decades ago. The success 
of (perturbative) QFT is adequately borne out by its accurate predictions of 
the cross sections for particle scattering. The first experiments on particle 
scattering was carried out by Rutherford, in 1911, who studied the scattering 
of particles from thin foils. Ever since, scattering experiments have been a major 
tool in our understanding of the constituents of matter.
Such experiments have put on a firm basis the standard model of particle physics, 
which, till date is the most successful theory of elementary particle interactions. 

Scattering of elementary particles can be considered in various limits of 
energy. One can, for eg., consider deep inelastic scattering, with a large transverse
momentum transfer, as was done in the HERA experiments, and compare the particle 
cross sections so obtained with theoretical predictions. It turns out that in
certain energy regimes, the scattering process becomes extremely simplified, and
several beautiful predictions can be made about the same. These regimes may not
always be accessible with present day experimental devices, but nevertheless,
are extremely important for furthering the theoretical understanding of QFT, 
and one can always hope that some future experiment will be able to test these
theoretical predictions. Such an energy regime, which offers great simplification
to the usual QFT processes is the almost forward scattering regime. 

Consider, for eg. the almost forward scattering of two electrons in four dimensional
space-time. Let us take the two electrons to be moving along the $z$ direction. 
Calling (a combination of the) $z$ and $t$ directions as the longitudinal 
directions, and the $x$ and $y$ directions as the transverse ones, one is dealing 
with extremely high longitudinal momenta. Further, in this type of scattering 
processes, we consider the transverse momentum transfer of the electrons to 
be extremely small compared to the longitudinal momenta, and also consider the
impact parameter to be large. There are several ways to deal with the computation 
of scattering amplitudes in this scenario. First of all, since the scattering is
almost forward, we may hope to employ semiclassical methods to derive the 
amplitude. This will consist of using the variables of the classical motion of the
electrons to construct their quantum mechanical wave functions and hence we can
read off the scattering amplitude. 

An alternative way of treating this problem is in the framework of quantum 
electrodynamics (QED). We can calculate the amplitude using the Feynman diagrams 
for electron-electron scattering and at the level of the diagrams make the 
approximations corresponding to the high energy limit. We will elaborate this in
great detail in the following sections, but let us mention at this point that
using the diagrammatic rules of QFT, one can exactly calculate this amplitude,
and the result turns out to agree exactly with the semi-classical
calculation. 

A third way of treating the same problem is to impose the said simplifications
at the level of the action itself. Namely, since the longitudinal and the 
transverse directions behave in vastly different ways, we can associate two 
different length scales to the two, and simplify the usual QED action. Again, as
we will discuss later, this gives the same result as the other two approaches. 

The fact that we could treat the same problem from three different points of 
view clearly demonstrates the simplification of the problem in this high energy 
regime. It is one of the few cases where a QFT amplitude can be constructed
fully, to all orders in the coupling constant. Hence, a study of such amplitudes
is important and interesting from the theoretical point of view, with the said
caveat that it is difficult to probe this from the point of view of present
day experiments. Indeed, one can try to probe experimentally viable energy 
scales by considering corrections to the exact formulae in the high energy case.
This has been an important topic of research in the past few decades, but will
not concern us here.

In this paper, we will be more interested in studying the regimes of 
ultra-high scattering for a different theory, namely non-commutative QED (NCQED).
This is interesting for the following reasons. A study of forward scattering
in QED shows a remarkable pattern for the amplitude. We can calculate all-orders
scattering amplitudes, and it can be seen that whereas the amplitude upto 
the sixth order is free from logarithmic divergences (arising out of terms like
log$s$, where $s$ is the centre of mass energy), such divergences appear in 
the eighth order of the coupling constant. As a consequence, the photon 
propagator in QED is not reggeized. A very different thing happens in the
context of high-energy quark-quark scattering in QCD. Here, the gluon propagator
can be replaced by an effective reggeized propagator, and the amplitude can
be understood in terms of multi-reggeon exchange. This is a consequence
of the fact that in each order of perturbation theory in QCD, there are 
logarithmic factors of the centre of mass energy. 

One may suspect these vastly different behaviours to arise due to the fact that
whereas in QED, the photon does not couple to itself, in QCD, the gluon does.
Now, NCQED is known to have three and four photon 
couplings. Hence, we can try to study the behaviour of amplitudes in NCQED
and understand its similarities and differences with the QED and QCD 
calculations. 

It is this issue that we address in this paper. To begin with, we study the 
problem of almost forward scattering of electrons in QED. We then study the same 
problem in NCQED. We find that the singularity structure of the scattering process 
in NCQED starts differing from the sixth order in the coupling constant, and is 
indeed different from the QED case. 

The paper is organised as follows. In section 2, we develop the basic tools for 
studying the scattering process, using semi-classical quantum mechanics, high energy
approximations to Feynman diagrams, and scaling arguments. In section 3, we study the 
same process, now from the point of view of NCQED, and present the results of the
Feynman diagram calculations for this process in NCQED. Section 4 ends with
some discussions and conclusions.

\section{Semi-Classical Calculations}

In this section, we deal with the problem of high energy scattering of two
charged scalars. We will be ultimately interested in promoting our results to
the case of fermions, and this will just involve the multiplication of the 
scalar result by an appropriate kinematic factor \cite{jko}.

\subsection{The Shock Wave Picture in QED}

The scenario we will consider has already been spelt out in the introduction.
Namely, we take two charged scalar particles moving with a very high value of
the centre of mass energy, $s$ and we will be interested in the limit
when $s~\to \infty$. It is convenient to Lorentz transform to the frame of
one of the particles, which is then at rest w.r.t this frame. Hence, we 
consider an observer sitting on one of the particles and
making observations on the second, which moves along the $z$ axis with a 
velocity $v\simeq c$. 

We can calculate the electromagnetic field due to the moving particle 
by considering the particle at rest and then Lorentz boosting it to (almost)
the speed of light. Let us first understand this qualitatively. The electric
field of the particle at rest is spherically symmetric. As the particle
starts to move along the $z$ axis, this spherical symmetry is lost, and the 
field lines tend to get concentrated away from the direction of motion, i.e 
on the $x-y$ plane. When the particle approaches the speed of light, all
the field lines are concentrated on this plane, and the component of the
electric field along the $z$ axis is zero, $E_z=0$. Now consider the test 
particle, which is at rest. An observer residing on this particle will not
feel the electric field due to the other since the latter is concentrated only
in the $x-y$ direction. It will feel the effect instantaneously, when the
moving particle passes directly overhead. Thus the effect will be
instantaneous, and such an interaction is commonly known as the shock wave
interaction. Consider a particle at rest, which has the vector potential at
a distance $r$ given by
\be
A^{\mu}=\left(\frac{e}{r},0,0,0\right)
\ee
Where $r=\sqrt{x^2+y^2+z^2}$. 
Using Lorentz transformations, we can construct the boosted form of the 
potential, and taking the limit of the velocity to approach the speed of light,
it can be shown that the electric and magnetic fields take the form \cite{jko} 
\bea
E^z&=&0~~~~~~~~~~E^i=\frac{er^i_{\perp}}{2\pi r_{\perp}^2}\delta(t-z)
\nonumber\\
B^z&=&0~~~~~~~~~~B^i=-\frac{e\epsilon^{ij}r^j_{\perp}}{2\pi r_{\perp}^2}\delta(t-z) 
\label{shockwave}
\eea
Now, we can construct the gauge potentials that give rise to the above
fields. These can be shown to be
\be
A^0=A^z=-\frac{e}{2\pi}{\mbox{ln}}\left(\mu r_{\perp}\right)
\delta(t-z),~~~~~{\bf A}^{\perp}=0
\label{pots}
\ee
Since the interaction is instantaneous, the wave
function of the free particle is a plane wave before the interaction. After
the interaction, this is modified, and remembering that the gauge interaction
changes the ordinary derivative to the covariant derivative, we can calculate
the change in the wave function which will be a factor of the form 
$e^{ie\int A^{\mu}dx^{\mu}}$. Carrying out the calculation using the explicit
form of the potential as in (\ref{pots}), the final wave function
is of the form
\be
\psi = {\mbox{exp}}\left(-i\frac{ee'}{4\pi}{\mbox{ln}}\mu^2r_{\perp}^2\right)
\psi_0
\ee
where $\psi_0$ is a free wave function (related to the initial free wave by
continuity conditions). 

From this, we can read off the scattering amplitude by expanding the final
wave function in the basis of plane waves, and the result due to Jackiw
et. al. \cite{jko} is
\be
f(s,t)=\frac{1}{4\pi i\mu^2}\frac{\Gamma(1+i\alpha)}{\Gamma(1-i\alpha)}
\left(\frac{4\mu^2}{-t}\right)^{1-i\alpha}
\label{jkoamp}
\ee
where $s$ and $t$ are the usual Mandelstam variables, and $\alpha=
\frac{ee'}{4\pi}$.  
This is the well known amplitude for eikonal scattering of two scalar particles. In 
the case of fermions, the result in eq. (\ref{jkoamp}) gets multiplied
by a factor of $\frac{s}{2m^2}$.

\subsection{Feynman Diagram Approximation}

\bfi
\begin{center}
\mbox{\epsfig{file=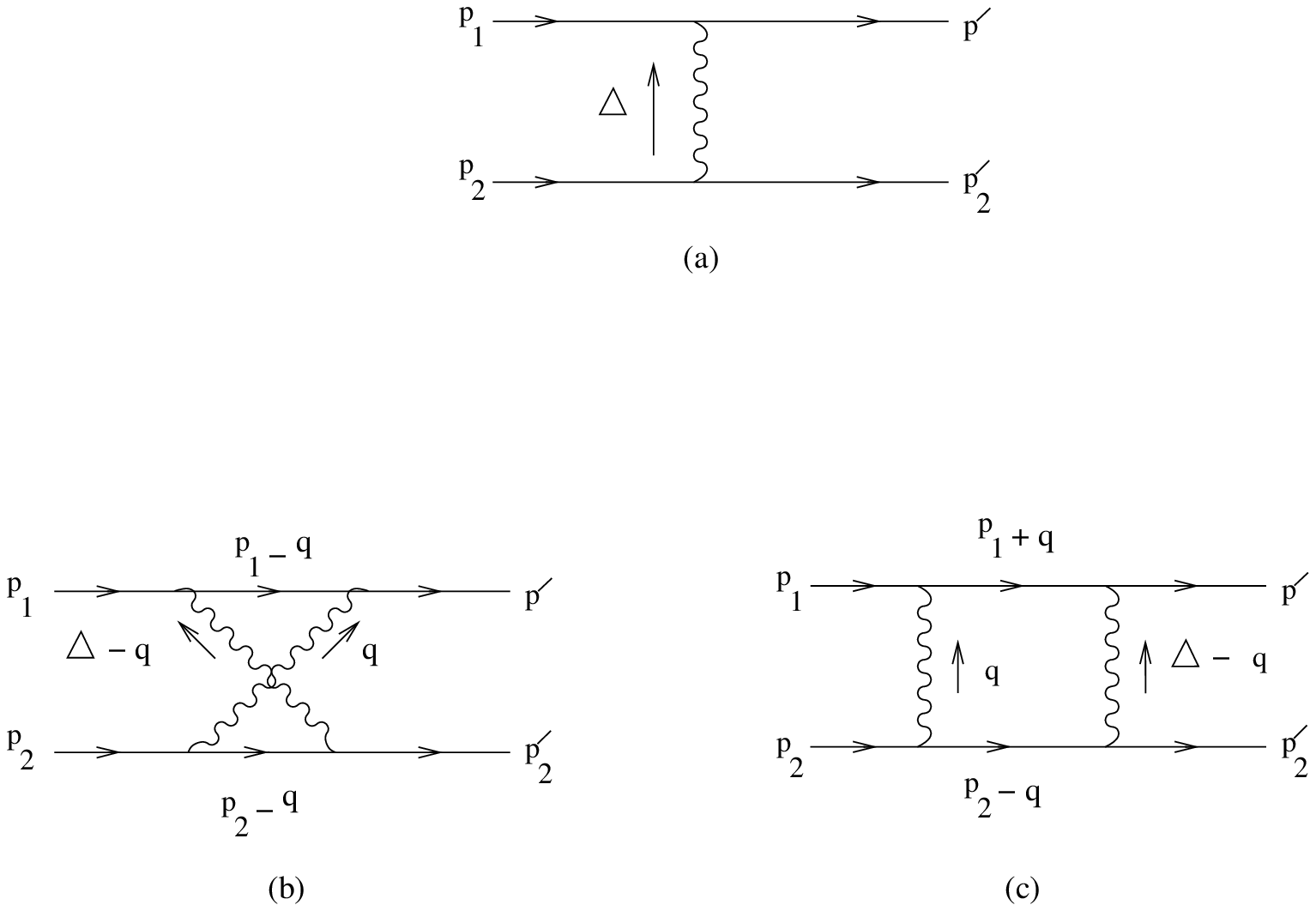,width=10truecm,angle=0}}
\caption{1 and 2-photon diagrams contributing to the semi classical amplitude in
e-e scattering in QED.}
\label{ladder1}
\end{center}
\efi

The same result as presented above can be succinctly derived from a Feynman diagram 
approach \cite{chengwu},\cite{itabar}. This consists of making the necessary 
high-energy approximations to Feynman graphs, in which, in the eikonal approximation, 
only ladder type diagrams contribute. These are shown for the cases of $1$, $2$, and 
$3$-photon exchange in figs. (\ref{ladder1}), (\ref{ladder2}). 

\bfi
\begin{center}
\mbox{\epsfig{file=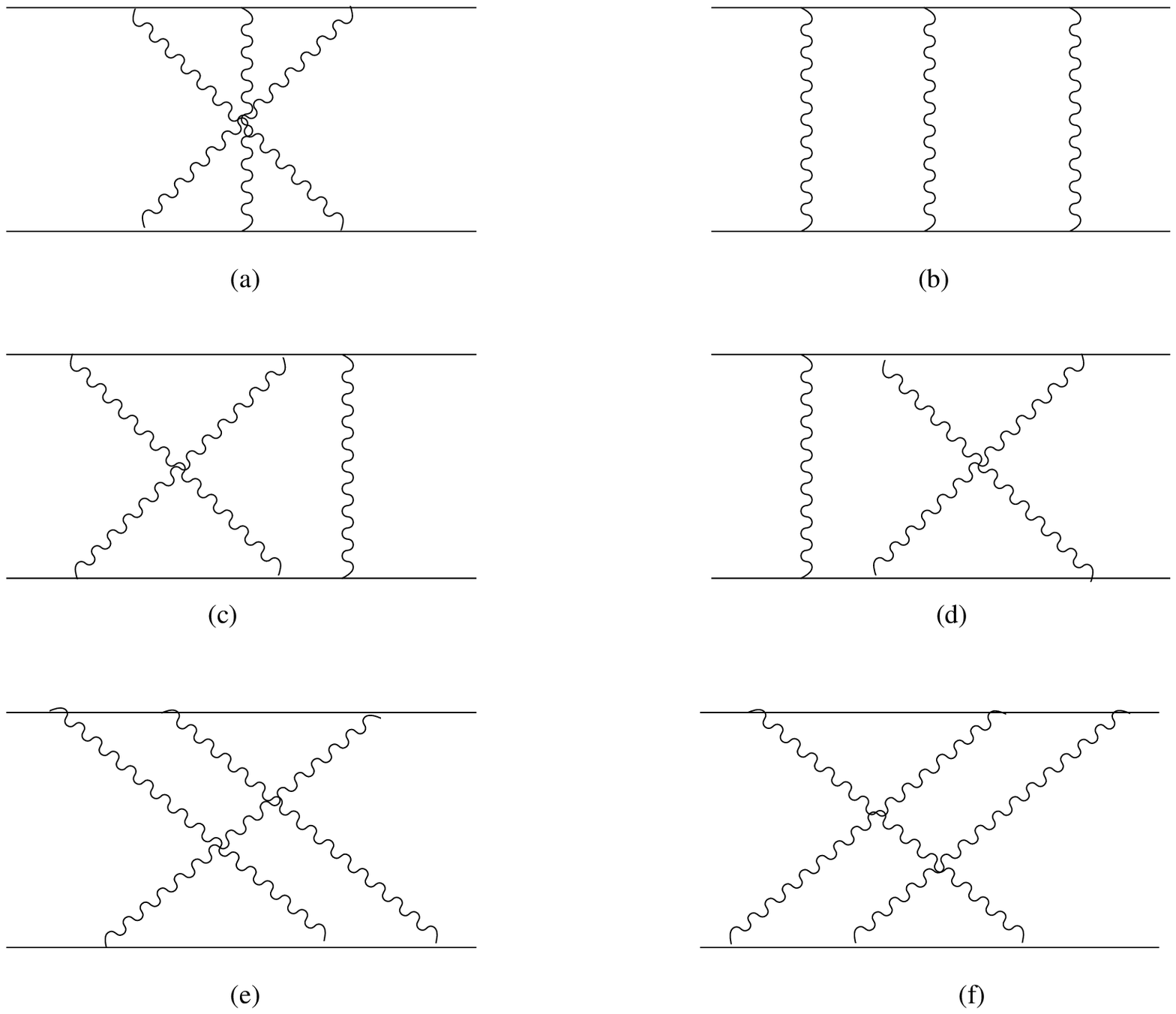,width=9truecm,angle=0}}
\caption{3-photon diagrams contributing to the semi classical amplitude in
e-e scattering in QED.}
\label{ladder2}
\end{center}
\efi

Consider, for example, the one-photon exchange amplitude. This is simply 
given by
\be
{\mathcal{M}}_1=\frac{e^2s}{2m^2}\frac{1}{\d^2+\mu^2}
\label{onephoton}
\ee
where $e$ is the electron charge, $m$ its mass, and $\d$ the transverse
momentum transfer. A fictitious mass for the photon is included to 
avoid infra-red divergences. We will assume this mass in all subsequent
formulae. Also note that the momentum transfer $\d$ is assumed to have
only transverse components, in line with the limit in which we are working. 
In Fourier space, (\ref{onephoton}) can be written as
\be
{\mathcal{M}}_1=-\frac{s}{2m^2}\int d^2x_{\perp}{\mbox{exp}}\left(i\d .{\bf
x_{\perp}}\right)\left[\frac{e^2K_0\left(\mu x\right)}{2\pi}\right]
\ee
where $x$ is $|{\bf x}_{\perp}|$ and $K_0$ is the modified Bessel function.
Similarly, one can calculate the amplitude corresponding to figure 
(\ref{ladder2}) and the result turns out to be
\be
{\mathcal{M}_2}=\frac{is}{4m^2}\int d^2x_{\perp}{\mbox{exp}}\left(i\d .{\bf
x_{\perp}}\right)\left[\frac{e^2K_0\left(\mu x\right)}{2\pi}\right]^2
\ee
This suggests a beautiful pattern for the general $n$ photon exchange 
amplitude, and we quote the result here \cite{chengwu},
\be
{\mathcal{M}_n}=-\frac{is}{2m^2n!}\int d^2x_{\perp}{\mbox{exp}}\left(i\d .{\bf
x_{\perp}}\right)\left[
-\frac{ie^2K_0\left(\mu x\right)}{2\pi}\right]^n
\ee

And the asymptotic form of the sum of amplitudes with $n$ photon
exchange is given by the series
\be
\sum_{n=1}^{\infty}{\mathcal M}=\frac{is}{2m^2}
\int d^2x_{\perp}{\mbox{e}}^{i{\bf \Delta}.{\bf x_{\perp}}}
\left[1-{\mbox{exp}}\left(-\frac{ig^2K_0(\mu x)}{2\pi}\right)
\right]
\ee
The integration can be carried out explicitly. Note that since $\mu$ is small, 
the Bessel function can be expressed as $-\left({\mbox{ln}}\mu x\right)$, and 
thereafter the integration is standard. The final result is exactly the same as 
in eq.(\ref{jkoamp})

We will make a few comments here. First of all, note that
the integration is explicitly over the transverse directions only. 
\footnote {The longitudinal momenta contribute delta functions \cite{chengwu}
and the final expression for the amplitude involves only transverse momenta.}
Hence, in the semi-classical
limit, the problem can be thought of as a two-dimensional one. Further, the 
log that appears in the small argument expression for the Bessel function 
is strongly reminiscent of the free two dimensional propagator. Therefore,
in some sense, one expects that QED in this extreme high energy limit should
be described by a free two dimensional theory. In the next subsection, we show that
this is indeed the case. 

In order to reduce the theory to an effective 
two-dimensional theory, we need to make certain approximations at the level of
the action, which is now commonly known as the Verlindes' scaling argument.
Let us therefore proceed to study this. 

\subsection{Verlindes' Scaling Argument}

\footnote{To the best of our knowledge,
the treatment in this subsection has not appeared elsewhere, but for a 
closely related discussion, see \cite{maggio}.}
An elegant way of treating high energy scattering is to make simplifications
at the level of the action of the theory under considerations \cite{vv}.
Noting that the square roots of the Mandelstam variables $s$ and $t$ measure 
the typical momenta associated with the longitudinal and the transverse directions,
it is reasonable to associate $two$ length scales, along these
two directions, the characteristic length scale along the
longitudinal direction being much smaller than that along
the transverse direction. Hence we can expect a simplification of the
action by performing a rescaling of the coordinates,
\begin{eqnarray}
x^{\alpha}~& \rightarrow&~ \lambda x^{\alpha} \\
x^{i}~ &\rightarrow&~ x^{i}
\end{eqnarray}
where $\alpha~ ,~ \beta$ runs over the light cone
coordinates $+,~ -$ and $i$ signifies the space coordinates
$x,~ y$, and $\lambda$ is a scale parameter that we
will take to be very small.\footnote{An equivalent statement would
be to scale the components of the metric \cite{vvgrav}.}

The gauge potentials $A^{\mu}$ have the dimension of $(length)^{-1}$ and their 
transformation property under the rescaling is given by
\begin{equation}
A^{\alpha}~ \rightarrow~ \lambda^{-1}A^{\alpha},
~~~~~A^{i}~ \rightarrow~ A^{i}.
\end{equation}
Performing this scaling in the action, and taking the limit $\lambda \to 0$,
we obtain the reduced action
\begin{equation}
S~=~-{1 \over 4}\int d^{4}x \left(\lambda^{-2}F_{\alpha \beta}
F^{\alpha \beta}~+~2F_{\alpha i}F^{\alpha i}\right)
\label{reduced}
\end{equation}
Clearly, in the above action, the dominant contribution comes from 
configurations with $F_{\alpha\beta}=0$, which can be translated into
$E_z=0$, thus making contact with the shock wave in eq.(\ref{shockwave}).
This implies the pure gauge condition, $A_{\alpha}=\partial\Omega$, and
hence the action (\ref{reduced}) can be written as
\be
F_{\alpha i}~=~\left[\partial_\alpha  \left(A_{i}~
-~\partial_{i} A_\alpha \right) \right]
\label{reduced1}
\ee
The Euler-Lagrange equation for $A_i$ calculated from this action implies
that $(A_i-\partial_i\Omega)$ is a harmonic function, and putting this
information back into the action, we obtain a semi-classical value of the
action,
\be
S~=~-{1 \over 4}\int d^{2}z\int_{- \infty}^{\infty}
dx^{+}\partial_{+} \left(A_{i}~-~\partial_{i}\Omega \right)
\int_{- \infty}^{\infty}dx^{-} \partial_{-} \left(A_{i}~-~
\partial_{i}\Omega \right)
\label{ac1}
\ee
From the above equation, it would seem that under a redefinition of
our gauge fields via a gauge transformation, we could, in fact, get rid
of all local interactions. The important subtlety here is that \cite{vv} 
arbitrary gauge transformations, as suggested by (\ref{ac1}), are not
allowed at infinity, and that the asymptotic values $\Omega$ should
survive. We impose the condition that the transverse gauge field $A_i$ take
the same values at the two ends of each Wilson line. Then, denoting the
asymptotic values of the field $\Omega$ at the end points of the Wilson lines
to be $g_1,g_2$ and $h_1,h_2$, and calling $g_1-g_2=g$ and $h_1-h_2=h$, we
finally obtain the action
\be
S~=~-{1 \over 4}\int d^{2}z~\partial_{i}g \partial_{i}h
\ee
This is thus the free field action for the fields $g$ and $h$ which have dynamics
only in the transverse directions, their values in the longitudinal direction
being fixed at the asymptotic infinities of the Wilson lines. 

It can be shown \cite{nachtmann} that the scattering amplitude of two electrons
in this high energy limit can be approximated by the expression
\be
f(s,t)=\frac{is}{2m^2}\int d^2z {\mbox{e}}^{iq.z}\left<V_+(z)V_-(0)\right>
\label{amp}
\ee
where $V_{\pm}$ are the Wilson lines defined by
\be
V_{\pm}~=~{\mbox{exp}}\left(e\int_{- \infty}^{\infty} dx^{\pm}
A_{\pm}(z) \right)
\label{wilson}
\ee
$e$ denoting the charge of the particles. Carrying out the integrations 
explicitly, we see that the scattering amplitude reduces to
\be
f(s,t)~=~\frac{is}{2m^2}\int d^{2}z {\mbox{e}}^{iq.z}~\left<{\mbox{e}}^{g}~
{\mbox{e}}^{h}\right>
\ee
Now, using $\left<{\mbox{e}}^g{\mbox{e}}^h\right>={\mbox{e}}^{\left<gh\right>}$,
and noting that the propagator in two dimensions is a log, we obtain
\be
f(s,t)~=~\frac{is}{2m^2}\int d^{2}z{\mbox{e}}^{iq.z}~{\mbox{e}}^
{-\frac{ie^2}{4\pi}ln|z|}
\ee
The integration can be performed in the same way as in the last subsection and
again yields the same result as in eq.(\ref{jkoamp}).

Let us briefly summarise what we have obtained till now. We have described 
three different ways of calculating the ultra-high energy scattering amplitude
for two electrons. The first method involved a semi-classical calculation, 
where one of the electrons was boosted to the speed of light and the change
in the wave function of the other was calculated (in the frame in which the
second electron is at rest). The second method involved approximations in the
Feynman diagrams of QED, and the third involved truncating the full QED action
in order to obtain an effective two-dimensional free field theory. 

Before proceeding further, let us point out that the shock wave picture has been 
studied in great details in the context of gravitational Planckian scattering. See, 
for eg. \cite{thooft},\cite{dasmajum}.

We are now ready to discuss the same scenario in the context of non-commutative
QED (NCQED), i.e QED in non-commutative spaces. This has received a lot of attention 
of late \cite{ncqed}, following the motivation of such spaces from the point of view 
of string theory \cite{sw}. We will be interested in calculating the scattering 
amplitudes of two electrons in non-commutative space. Our main motivation for
doing so is the formal similarity between \nc and QCD. It is well known, that
\nc has three and four-photon interactions, which are absent in usual QED. One
might, therefore, be tempted to think that the high-energy scattering 
behaviour of \nc would in some sense, be similar to that of QCD. The latter has
been well studied in \cite{chengwu} and one of the main results therein is the
appearance of logarithmic factors of the centre of mass energy in the
scattering amplitude, which are absent in QED upto sixth order in the coupling. 
Further, it has been shown that in this energy regime, the gluon exhibits
Regge behaviour, while the photon doesnt. We would like to understand the same
phenomena in the context of \nc, which is our topic of study in the next section. 

\section{High Energy Scattering in Non-Commutative QED}

We now address the question of high-energy scattering in NCQED. Our hope here 
is that there would be some features which are qualitatively different 
from those in usual QED, because of the widely different natures of 
these two theories (the Feynman rules for NCQED are presented in 
the appendix). For earlier work on scattering in non-commutative theories,
see, for eg., \cite{rivelles}.  

\subsection{The Fate of The Shock Wave}

First of all, we would like to understand the nature of the ``shock wave'' in 
\nc. Here, we run into an obvious problem. Due to the lack of Lorentz invariance
in the theory, the results for the scattering amplitude are expected to be 
frame dependent, and although, one could, in principle, boost one of the charges,
the semiclassical approach is expected to be difficult in this case, due to
the non-linearity of the noncommutative theory and indeed, the shock wave picture
ceases to be valid, as will be made clearer in
a while. Note that we could, however, try and apply the other two procedures that
we have outlined. Namely, we could make approximations at the level of Feynman
diagrams for \nc, and we could scale the action, as is appropriate for such
high energy scattering. Let us study the second procedure first. We will argue
non-commutative corrections invalidate the shock wave picture.  

Let us begin by noting that the non-commutative generalisation of the free
Maxwell action is to replace the usual product by the star product and 
expressing the field strength in terms of the gauge potential $\amuhat$.
\begin{eqnarray}
{\mathcal L}&=&-\frac{1}{4}{\hat F_{\mu\nu}}\star{\hat F^{\mu\nu}}\nonumber\\
{\hat F_{\mu\nu}}&=&\partial_{\mu}\anuhat - \partial_{\nu}\amuhat
-ie\left(\amuhat\star\anuhat - \anuhat\star\amuhat\right) 
\label{ncmaxwell}
\end{eqnarray}
The star product here is defined by
\be
\left(\Phi_1\star\Phi_2\right)(x)=\left[{\mbox{exp}}\left(\frac{i}{2}
\theta^{\alpha\beta}\partial_{\alpha}\partial'_{\beta}\right)
\Phi_1(x)\Phi_2(x')\right]_{x=x'}
\ee
and we define the Moyal bracket as 
\be
\left[A_{\mu},A_{\nu}\right]_{MB}=A_{\mu}\star A_{\nu}-A_{\nu}\star A_{\mu}
\ee
Using the Seiberg-Witten map that expresses the non-commutative gauge fields and
gauge parameters in terms of the commutative ones order by order in 
$\theta$, \cite{sw}, the Lagrangian of the theory (upto first order in $\theta$), 
is given by \cite{bichl}
\be
\mathcal{L} = -\frac{1}{4}F_{\mu\nu}F^{\mu\nu} + \frac{1}{8}
\theta^{\alpha\beta}F_{\alpha\beta}F_{\mu\nu}F^{\mu\nu} 
-\frac{1}{2}\theta^{\alpha\beta}F_{\mu\alpha}F_{\nu\beta}F^{\mu\nu}
+O\left(\theta^2\right)
\ee
Let us consider very high energy scattering in \nc  using this
Lagrangian. We assume, without loss of generality that the components of
$\theta$ are non-vanishing only along the transverse directions, i.e
$\theta^{12}=-\theta^{21}=\theta$. It has been shown that for time like 
non-commutativity, the theory is often non-unitary \cite{gomis} and we will
not be concerned with such aspects for the purpose of this paper. 

Note that $\theta$ has the dimensions of the square of the
transverse length and its scale is fixed accordingly.  
In particular, following \cite{vv}, we set the length scale
of the longitudinal directions to be $\lambda << 1$ and that of the transverse
dimensions to unity. Hence, in the action, $\theta$ is not scaled. 
We are, in a sense,  making the simplifying assumption that $\theta$ does not 
set a new scale in the theory. In what follows, we will show that even with
this limitation, it is clear that the usual shock wave picture of high-energy
scattering in QED is invalidated. Scaling the action as in \cite{vv},
we obtain the leading term in the Lagrangian (equivalent to the first term
in eq.(\ref{reduced}) as 
\be
\frac{1}{\lambda^4}F^{\alpha\beta}\left[-{1\over 4}F_{\alpha\beta}
+\frac{1}{8}\theta^{ij}F_{ij}F_{\alpha\beta}F^{\alpha\beta}
-{1 \over 2}\theta^{ij}F_{\alpha i}F_{\beta j}F^{\alpha\beta}\right]
\ee
Expressing the field strength in terms of the electric and magnetic
fields in the usual way, we find that the dominant contribution to the
path integral comes from configurations in which 
\be
E_z=-\theta^{12}\left(E_xB_x + E_yB_y\right)
\label{nceikonal}
\ee
This result, which is valid in any frame of reference in which the longitudinal
momentum is very large compared to the transverse ones (in particular the 
centre of mass frame) is the \nc analogue of the shock wave picture of the last 
section.  From eq.(\ref{nceikonal}), it is clear that the interaction between two 
electrons moving at ultra-high energies is no longer instantaneous in this
framework. The non-vanishing $E_z$ causes the two particles to interact
at all times, and hence the simple calculations of the last section can
no longer be performed here. 

The same result can be obtained by starting with the pure
non-commutative Maxwell action, eq. (\ref{ncmaxwell}), and 
noting that in terms of this action, the scaling argument produces
the pure gauge condition in \nc, ${\hat F}_{\alpha\beta}=0$ where 
\be
{\hat F}_{\alpha\beta}=\partial_{\alpha}A_{\beta}-\partial_{\beta}A_{\alpha}
-ie\left[A_{\alpha},A_{\beta}\right]_{MB}
\ee
and on the r.h.s we have the Moyal bracket of the gauge fields. Expanding the r.h.s to 
leading order in $\theta$, this equation gives the same condition as eq.(\ref{nceikonal}).

It is difficult, in this framework, to estimate the corrections to the
usual QED amplitudes for \nc. As we have already seen, the shock-wave picture
is not valid in this framework (at least in the centre of mass frame) and 
an effective way to calculate the \nc corrections to the shock
wave using semi-classical techniques is not known to the present authors. 
The problem arises because of the fact that the dispersion relations in \nc
are non-linear \cite{jackiwnc} and it is difficult to solve this set of Maxwell's 
equations explicitly. 

Therefore, we are led naturally to the approach (2) described in the last 
section. Namely, we would make approximations in the Feynman diagrams for 
\nc in the high energy limit, in order to evaluate the corrections. Naively,
it might seem that these corrections will be negligible, since we have tuned
the non-commutativity parameter $\theta$ to take values only in the 
transverse directions. As we will now show, this is not true. It turns
out that there are in fact important leading log corrections to the
high-energy scattering amplitude in \nc.

A simple way to understand this is as follows. Consider the usual QED. It is
well known \cite{chengwu} that in this theory, there are no leading log 
corrections to electron-electron scattering upto the sixth order. At the eighth
order, such corrections start appearing as a result of uncancelled 
logarithms arising due to electron loops inside ladder diagrams. The cancellation
of the leading logs (upto sixth order) can be seen most easily by the method
of ``flow diagrams.'' This technique was invented in \cite{chengwu} in order
to keep track of the factors that contains the logarithms of the centre of mass
energy in the high-energy limit. It turns out that in QED, although all the
individual diagrams have such logarithmic factors, these are cancelled when
one sums all diagrams, and hence the absence of leading logs upto sixth order.

The logarithmic factors appear in the individual QED diagrams, as an 
infrared cutoff to the longitudinal momenta. For NCQED, since we have chosen the 
directions of noncommutativity to be along the transverse directions, these logarithms
will appear as in usual QED. However, we expect the scenario to be modified 
in the context of \nc because of two reasons. Firstly, the Feynman rules for \nc 
contains extra phases, (which, in our case depend only on transverse momenta), and
which might be different for different diagrams. Hence the leading logs might
not cancel. The second reason for which we might expect leading logs in \nc 
amplitudes is because of the QCD like processes involving three and four
photon interactions (fig. (\ref{qcdlike}). Let us therefore proceed to study 
these in some details.

\subsection{Feynman Diagram Calculations for NCQED}

To be specific, we will, in the spirit of our QED discussion, consider only 
multi-photon exchange diagrams of QED and the QCD-like corrections to the same in 
\nc, upto the sixth order, as in figures (\ref{ladder1}) and (\ref{ladder2}). 
This will bring us to the issue of the leading corrections to the eikonal picture 
that we have described in section 2. 

Of course, this set of diagrams is not exhaustive, and there are other diagrams to the 
same orders that we are considering, like radiative corrections to the form factors, etc. 
These diagrams will be important in modifying the vertex functions and the photon 
propagator \cite{chengwu}. \footnote{A discussion on such diagrams upto the sixth order 
in the coupling constant can be found in chapter 11 of the book \cite{chengwu}.} 
These will bring in the important issues of renormalisation and the UV/IR mixing 
phenomena in non-commutative gauge theories \cite{mst},\cite{mvs}. For the purpose 
of this paper, we do not consider these diagrams further, and assume that as in
the case of usual QED, the 
behaviour of these in the high energy limit will not modify our results on the 
logarithmic factors of the centre of mass energy. As we have already pointed
out, the logarithmic factors of $s$ appears in our calculations as infrared cutoffs 
for the longitudinal momenta (see Appendix) and the non-commutative parameter 
$\theta$ will be set to zero along these directions. Hence, in a sense, we expect 
these factors of ${\mbox{log}}(s)$ to remain unaffected in \nc. It will 
indeed be interesting to compute the radiative corrections to the fermion form 
factors in the high-energy limit in \nc, and we will leave this to a future work. 

\bfi
\begin{center}
\mbox{\epsfig{file=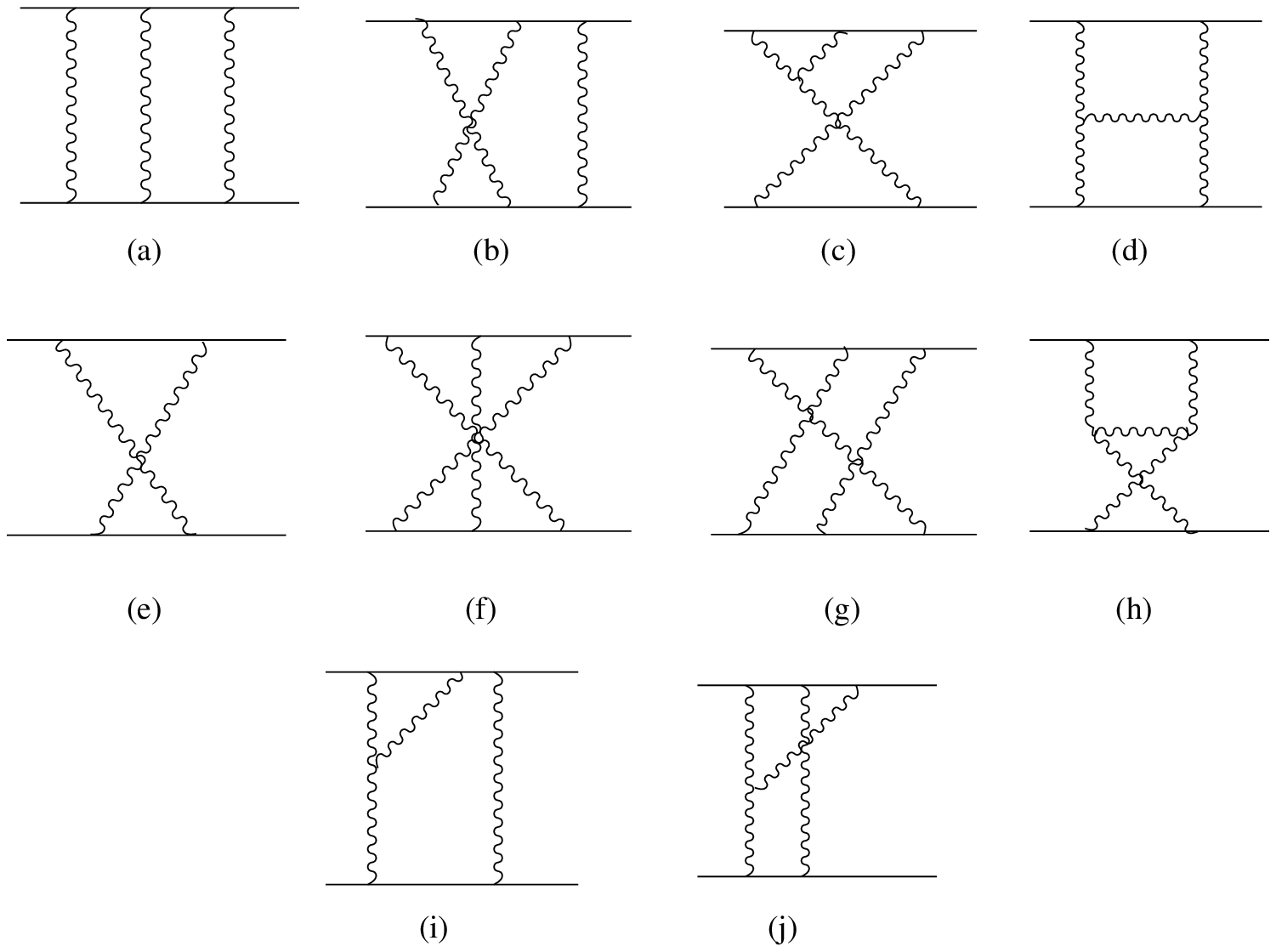,width=11truecm,angle=0}}
\caption{sixth order interaction diagrams that contribute in 
e-e scattering in NCQED.}
\label{qcdlike}
\end{center}
\efi

Let us first specify the regimes of our calculation. We will be working in the centre 
of mass frame, and will use the usual Mandelstam variables, $s$, $t$ and $u$. We will 
work in the almost forward regime, where $t << s$.  It turns out that the calculations 
become simple if we choose the $\theta$ regime to be
\be
\theta t = \theta\d^2 << 1
\ee
where $\theta$ is the non-commutativity parameter. Hence, we will be working
in the regime of small $\theta$.  

We will be using the method of flow diagrams as in \cite{chengwu} (see also
\cite{feng}). We will not
elaborate on this method here (a brief review is contained in the Appendix), 
but let us point out some salient features 
of this technique. It consists of properly choosing the poles of the propagators in
the Feynman diagrams, such that the loop integrals are simplified. In order to
achive this, one draws all possible flow diagrams indicating the possible
directions of the negative components of the loop momenta. Integration over the
positive components (in the light cone sense) is then carried out, using the 
residue theorem, and integration over the $(-)$ components of the loop momenta are 
seen to produce the log factors in the diagram.  

Let us now consider the second and fourth order diagrams in NCQED in the 
extreme high energy limit. Consider first the second order diagram, depicted
in fig (\ref{ladder1}). In the centre of mass frame, the incoming and 
outgoing momenta have only longitudinal components, whereas the parameter
$\theta$ has only transverse components in our scenario. Hence, the phase
factor can be seen to be unity. Let us now consider the fourth order 
diagrams. There are two of them, as in fig. (\ref{ladder1}), and each of them
has one flow diagram, which is the same as the labelling of momenta shown. 
Using the same arguments as before, we find that in diagram (b) of fig. (\ref{ladder1}),
the net phase factor is ${\mbox{exp}}(i\qperp\times\d)$ while for diagram
(c) of fig. (\ref{ladder1}), the phase factor is unity. 
The logarithmic factors for the amplitudes for these diagrams, including the 
phase factors are respectively, 
\bea
{\mathcal M}_a=-\frac{e^4s{\mbox{ln}}s}{4\pi m^2}\int 
\frac{d^2q_{\perp}}{(2\pi)^2}\frac{{\mbox{e}}^{i\qperp\times\d}}
{\left({\bf q_{\perp}}^2\right)
\left[\left(\Delta-{\bf q_{\perp}}\right)^2\right]}~=~
-\frac{e^4s{\mbox{ln}}s}{4\pi m^2}I'\left(\d\right)
\nonumber\\
{\mathcal M}_b= \frac{e^4s{\mbox{ln}}s}{4\pi m^2}\int 
\frac{d^2q_{\perp}}{(2\pi)^2}\frac{1}
{\left({\bf q_{\perp}}^2\right)
\left[\left(\Delta-{\bf q_{\perp}}\right)^2\right]}~=~
\frac{e^4s{\mbox{ln}}s}{4\pi m^2}I\left(\d\right)
\eea
Where $I'$ and $I$ are the integrals with and without the phase factors
respectively. The term proportional to ${\mbox{ln}}s$ is now
\be
{\mathcal M}=-\frac{e^4s{\mbox{ln}}s}{4\pi m^2}\int 
\frac{d^2q_{\perp}}{(2\pi)^2}
\frac{{\mbox{cos}}\left(\qperp\times\d\right) -1}
{\left({\bf q_{\perp}}^2\right)
\left[\left(\Delta-{\bf q_{\perp}}\right)^2\right]}
\ee
and the usual QED amplitude, without the logs is now modified (under
a Fourier transform to position space) to
\be
{\mathcal M}=\frac{ie^4s{\mbox{ln}}s}{2\pi m^2}
\int d^2x{\mbox{e}}^{i\d.\xperp}K_0\left(\lambda|\xperp|\right)
K_0\left(\lambda|\xperp '|\right)
\ee
where $\xperp '=\sqrt{\xperp^2+\left(\theta\d\right)^2+\left(\xperp\times\d
\right)}$. It is interesting to note that one of the coordinates have been
shifted by a term proportional to $(\theta\d)$.
 
The $\theta~=~0$ case corresponds to these amplitudes in usual QED
\cite{chengwu}, and in this case, the integrals cancel each other. 

Now let us consider the sixth order amplitudes in NCQED. Some of the relevant
diagrams have been depicted in fig.(\ref{ladder2}). It is well known \cite{chengwu}
that Each of these diagrams have two flow diagrams, which we have presented
in the appendix, in figure (\ref{flow3}). Consider fig. (\ref{ladder2}). In order
to evaluate these diagrams, it is sufficient to calculate only two of them,
(b) and (d). This is because (b)  is related by $s \to u$ to (a). (c) and (d) 
line 2 are equal, and are related by $s\to u$ to those on line 3. 

The phase factors for these diagrams can be easily determined by the flow 
diagrams in fig. (\ref{flow3}). The important point to note here is that not
all the diagrams have similar phase factors. For eg. the diagrams of the middle
column have unit phase factors, and so have the lower two diagrams of the
last column. All the other diagrams have non-vanishing phase factors. 

A small comment is in order here. It is known that these diagrams often have
${\mbox{ln}}^2s$ terms, which are known to cancel in commutative QED. It can be
shown that the same happens in the case of NCQED too, and that there are no
such terms in the amplitude. We now list the expressions for the amplitudes
in figure (\ref{ladder2}). Using the notation
\be
I_1=\int\prod_{i=1}^3\frac{d^2 q_{i\perp}}{(2\pi)^2}{\mbox{cos}}\left(
\qoneperp\times\qtwoperp\right)
\frac{(2\pi)^2\delta^{(2)}\left(\sum_{i=1^3}{\bf q_{i\perp}}-\d\right)} 
{\prod_{i=1}^{3}\left({\bf q_{i\perp}}^2+\mu^2\right)}
\ee
\be
I_2={1\over 2}\int\prod_{i=1}^2\frac{d^2 q_{i\perp}}{(2\pi)^2}{\mbox{cos}}\left(
\qoneperp\times\qtwoperp\right)
\frac{{\mbox{ln}}~\frac{{\bf q}_{2\perp}^2}{{\bf q}_{1\perp}^2}\left(
{\bf q}_{2\perp}^2 + {\bf q}_{1\perp}^2\right)}{\left({\bf q}_{1\perp}^2\right)^2
\left({\bf q}_{2\perp}^2\right)^2\left({\bf q}_{1\perp}^2 - 
{\bf q}_{1\perp}^2\right)\left({\bf q}_{1\perp}+{\bf q}_{2\perp}+\d\right)^2}
\ee
and the values of these integrals for $\theta=0$ by $I_1'$ and $I_2'$ respectively,
the amplitudes are given by
\bea
{\mathcal M}_{a+b}&=& 
-\frac{e^6s{\mbox{ln}}s}{2\pi^2m^2}\left[I_2 + I_2'\right] ~~~~{\mbox{row 1}}
\nonumber\\
{\mathcal M}_{c+d}&=& 
\frac{e^6s{\mbox{ln}}s}{2\pi^2m^2}\left[I_2' + i\pi I_1'\right] ~~~~{\mbox{row 2}}
\nonumber\\
{\mathcal M}_{e+f}&=& 
\frac{e^6s{\mbox{ln}}s}{2\pi^2m^2}\left[I_2 - i\pi I_1\right] ~~~~{\mbox{row 3}}
\eea
For the case $\theta~=~0$, the sum of the above terms reduces to zero, signifying
that there are no logarithmic factors of energy in the sixth order e-e
scattering in QED. In this case, however, the sum is non-zero, and given by
\be
{\mathcal M}_6 =-\frac{ie^6s{\mbox{ln}}s}{2\pi m^2}
\int\prod_{i=1}^3\frac{d^2 q_{i\perp}}{(2\pi)^2}\left[{\mbox{cos}}\left(
\qoneperp\times \qtwoperp\right)-1\right]
\frac{(2\pi)^2\delta^{(2)}\left(\sum_{i=1^3}{\bf q_{i\perp}}-\d\right)}
{\prod_{i=1}^{3}\left({\bf q_{i\perp}}^2+\mu^2\right)}
\label{log1}
\ee
and hence there is a logarithmic contribution at sixth order, which is 
imaginary in nature. 

Now, we will consider QCD like corrections in NCQED. In fig. (\ref{qcdlike}),
we have drawn a few of the sixth order diagrams that will contribute to the
scattering process. The rest of these diagrams can be obtained by reflections
along a horizontal or a vertical line. For example, in fig. (\ref{qcdlike}),
diagrams (a), (b), (f), (g) and their permutations are the QED-like 
contributions of fig. (\ref{ladder2}). The remainder of the diagrams in
fig (\ref{qcdlike}) are the novel diagrams of \nc, and contribute to the
sixth order process. There are fifteen such diagrams, which can be obtained
as reflections of those presented here. For example, there are four diagrams
corresponding to (c) in fig. (\ref{qcdlike}), obtained from (c) by reflections
about a vertical or horizontal line. These are, of course all equal. 
  
Note that the four photon vertex in diagram (e) of fig. (\ref{qcdlike})
can be thought of as the fused diagrams corresponding to (d) and (h). We will
follow \cite{chengwu} and add the contribution of (e) to (d) and (h) before
we make the asymptotic calculation. The integrals in the expressions to follow
will be of two types.
\bea
\ti_1=\int\prod_{i=1}^{2}\frac{d^2 q_{i\perp}}{(2\pi)^2}{\mbox{sin}}^2
\left(\qoneperp\times \qtwoperp\right)
\frac{\qoneperp^2+\qtwoperp^2}{\prod_{i=1}^{2}
\qiperp^2\left(\d-\qiperp\right)^2}\nonumber\\
\ti_2=\int\prod_{i=1}^{2}\frac{d^2 q_{i\perp}}{(2\pi)^2}{\mbox{sin}}^2
\left(\qoneperp\times \qtwoperp\right)\frac{\d^2}{\prod_{i=1}^{2}
\qiperp^2\left(\d-\qiperp\right)^2}
\eea

The integral $\ti_1$ is an ultra-violet divergent integral. In usual
QCD, the terms proportional to $\ti_1$ cancel, when we arrange the diagrams
with the appropriate group theory factors, but this cancellation is not
seen here. It would be tempting to argue that these ultraviolet divergences
would disappear in the renormalised theory, but we do not have a concrete
proof of the same. We leave this issue for a future study, and for the 
moment we list the results for the QCD-like diagrams of fig. (\ref{qcdlike}). 
First, we list the terms that depend on $\ti_1$.
\bea
{\mathcal M}_{d+e+h}^{\ti_1} &=& 4\frac{e^6s}{8\pi^2m^2}\left[{\mbox{ln}}
\left(se^{i\pi}\right)\right]^2
\ti_1 -4\frac{e^6s}{8\pi^2m^2}\left[{\mbox{ln}}s\right]^2\ti_1 
\nonumber\\
{\mathcal M}_{j}^{\ti_1}&=&
-2\frac{e^6s}{8\pi^2m^2}\left[{\mbox{ln}}\left(se^{i\pi}\right)\right]^2
\ti_1+2\frac{e^6s}{8\pi^2m^2}\left[{\mbox{ln}}s\right]^2\ti_1\nonumber\\
{\mathcal M}_{c+i}^{\ti_1}&=&
-2\frac{e^6s}{8\pi^2m^2}\left[{\mbox{ln}}\left(se^{i\pi}\right)\right]^2
\ti_1-2\frac{e^6s}{8\pi^2m^2}\left[{\mbox{ln}}s\right]^2\ti_1
\nonumber\\
\eea
The contribution to these diagrams in the form of $I_2$ comes only from
diagrams (d),(e) and (h) and is given by the expression 
\be
{\mathcal M}_{d+e+h}^{\ti_2}=-i\frac{e^6s}{8\pi m^2}\left[{\mbox{ln}}s\right]\ti_2
\ee
The sum of the QCD like contributions is thus given by
\be
{\mathcal M}_{QCD-like}~=~-4\frac{e^6s}{8\pi^2m^2}\left({\mbox{ln}}s\right)^2\ti_1 
-\frac{ie^6}{8\pi m^2}\left(s{\mbox{ln}}s\right)\ti_2
\label{Qcdlike}
\ee
Note that the second term on the r.h.s of eq. (\ref{Qcdlike}) is formally 
similar to the expression in eq. (\ref{log1}), but in the former, the integrand
is proportional to $\d^2$, which we have assumed to be very small, and hence the 
${\mbox{log}}s$ contribution coming from (\ref{log1}) dominates. In the above 
calculations of the fourth and sixth order processes, we have neglected terms 
proportional to $\d\theta$ in the phase factors. These will provide subleading 
corrections to our results.
 
Our calculations above clearly show that for ultra-high energy scattering
in \nc, there are novel effects. The usual QED amplitude is corrected by
leading logarithms. Although these logarithms appear at the sixth order in
perturbation theory, it is possible that they will be significant in 
comparing with experimental predictions of \nc. 

\section{Discussions and Conclusions}

In this paper, we have studied almost-forward scattering in the context of 
\nc. We have shown that even in this regime where the kinematics make the
theory extremely simple, there is a difference between \nc and
usual QED. The difference arises expectedly because of the non-commutative nature 
of space, and even at the semi classical level, we have shown that the 
two theories exhibit completely different natures. Whereas in commutative 
QED, the shock wave picture can be effectively used to calculate scattering
amplitudes exactly, the same fails for the case of \nc. As we have
seen, the difference arose because in \nc, the interaction between two
fast moving particles become non-local in the longitudinal directions. 

Further, we have shown that in \nc, using Feynman diagram techniques to
calculate the amplitude perturbatively yields vastly different results from
QED, and that there appear logarithmic factors of energy in the amplitude
from the sixth order. In usual QED, these factors in fact cancel out 
when one adds the diagrams. For the QED like diagrams in \nc, they do not, since 
the diagrams come with different phase factors. Further, the QCD like
diagrams in \nc produce terms of the order of ${\mbox{ln}}^2s$.    

We have not considered the effects of renormalisation of \nc in this paper, and 
in particular the peculiarities like the UV/IR mixing encountered in the same. We 
have assumed, in a sense, that these issues will not modify our results on the leading 
logarithms. One way to see this is to note that in usual QED, the diagrams that are 
not of the ladder type are all of the order of $s$, and do not affect the leading 
log behaviour. It will be interesting to study this issue in some details, and we 
leave this to a future publication. 

Further, it would be interesting to see if there exists an effective
two dimensional theory of high energy scattering in \nc in lines with our 
discussion of section 2. One would be tempted to think that the asymptotic
behaviour of the amplitudes in \nc would be similar to those of QED but with
usual products replaced by star products. This is, in general, a complicated
problem to tackle, because many of the simplifying assumptions of QED 
will not be valid in this case, since different diagrams appear with different 
phase factors. However, it seems to us that even in \nc, the scattering 
amplitude can be calculated as the expectation value of two Wilson lines, 
as in section 2. It might be possible to calculate the logarithmic factors
presented in this paper from a two-dimensional field theory point of view. 

Also, it would be interesting to compute the corrections to our picture. We have 
made the simplifying assumption that $\theta$ is very small compared to the centre
of mass energy. It would be of interest to estimate the leading order correction terms. 
It is then that we hope to be able to connect to experimental results. We leave 
these issues to a future work.
\newpage
\vspace{2.0cm}
\noindent
{\bf Acknowledgements}

\noindent
It is a pleasure to thank G. Thompson for several helpful discussions and 
comments and for a careful reading of the manuscript. We would like to thank
Sumit Das for his comments on the manuscript. We thank N. Mahajan for helpful 
correspondence. We also thank L. Bonora, Venkat Pai, and A. Sen for discussions. 
T. S would like to thank T. Jayaraman who had introduced him to this subject.
Z. A would like to thank all the members of the ICTP high energy group for their
encouragement and support.

\newpage
\appendix
\noindent
{\bf \Large{Appendix}}
\vspace{1cm}

\noindent
In this appendix, first we briefly review the method of flow diagrams that we
have used in section 3. Then, we present the Feynman rules for \nc. Finally, we 
draw the flow diagrams for the sixth order QED-like diagrams for \nc. 
\section{Flow Diagrams}

The calculations of Feynman diagrams for high energy scattering are considered to
be difficult, but they are actually quite easy even for higher orders diagrams. 
The mathematical expression for a higher order diagram is of course long but 
because of high energy center of mass momenta approximations they can be 
calculated easily in a systematic way.  

A Feynman amplitude can be written
either over loop
momenta or we can introduce Feynman parameters and then perform the integration
over momenta variables. The asymptotic behaviour of Feynman amplitudes in the high 
energy limit is obtained by making approximations in either method. Historically, 
the high energy approximation was first made in Feynman parameter formulation. This 
method is rigorous but lengthy and intermediate steps are complicated but 
the final asymptotic expressions are quite simple. This shows that there is a 
relatively easy way to to get final results. The approximation in momenta 
variables achieves this. First of all, this approximation in momentum variables was
employed by Sudakov \cite{sudakov} and we also use momentum variables to study high
energy behaviour of scattering in gauge field theories.

We study the scattering process in center of mass frame in such a way that the 
spatial momenta of the two incident particles are along the positive z direction and 
negative z direction, with magnitude $w_1$ and $w_2$ respectively.
In order to calculate the Feynman amplitude for a given diagram, we first perform 
the integratin over the plus momenta explicitly by contour integration, 
and then the integration over minus momenta is performed by making suitable 
approximations. 

This later integration may be infinite
if we set $s$, (the square of the center of mass energy) to be infinite. 
This divergence is logarithmic.
In some diagrams we may find a divergence which is more than logarithmic but in gauge
field theories such divergence usually cancel when we sum all diagrams of a given 
order of perturbation. So the Feynman amplitude is proportional to $lns$ and the 
integrals over transverse momenta. There may be divergence from a single diagram arising
from this transverse momenta integration but again such divergences 
generally vanish when we sum over all diagrams of a given perturbation order.

\section{The Box Diagram}

\bfi
\begin{center}
\mbox{\epsfig{file=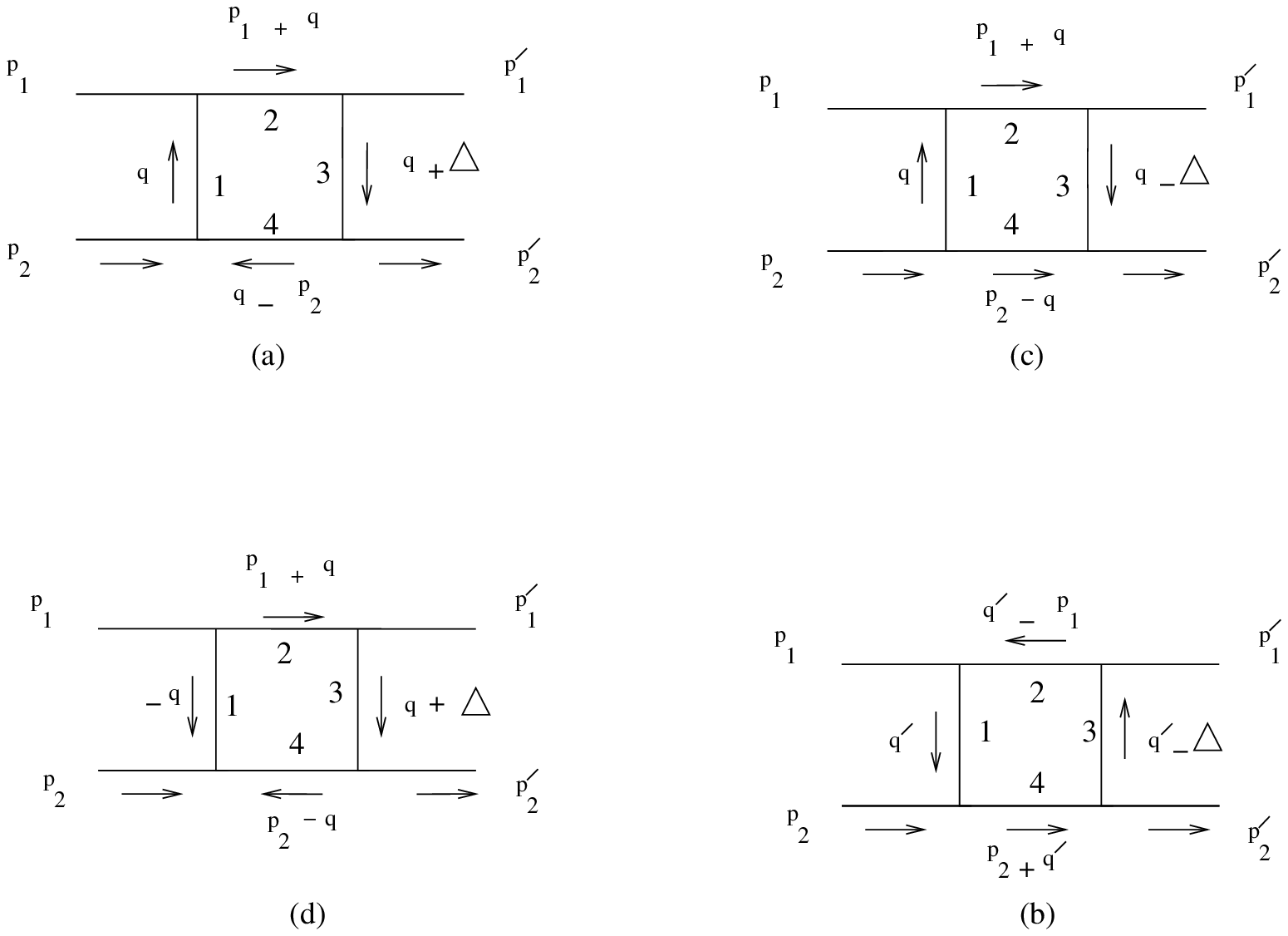,width=12truecm,angle=0}}
\caption{The flow diagrams corresponding to the box diagram (b) of fig. 
(\ref{ladder1}). Only diagram (c) contributes.}
\label{flow}
\end{center}
\efi
Let us illustrate this in some details. Consider the box diagram in fig. 
(\ref{flow}).
We assume, for simplicity, that the solid lines represent scalar particles. 
Take the propagator of momentum q. In light
coordinates we can write the denominator of this propagator as 
\be
q_+ q_- +q_\perp^2 +m^2 + i\epsilon
\ee
This propagator has one pole locatedd at
\be
q_+ =- \frac{q_\perp^2 +m^2}{q_-} -\frac{i\epsilon}{q_-}
\ee
This pole is located in lower half plane if $q_-$ is positive and if $q_-$ is 
negative, then this pole is in upper half plane. 
For a given Feynman diagaram there is no restriction on value of $q_-$ momenta,
it can be positive or negative. There are two different regions of kinematics 
belonging to the two different value of $q_-$ momentum. We represent this by the
two diagrams in fig. (\ref{flow}), (b) and (c). These are known as the flow 
diagrams. By definition minus momenta is positive
in a flow diagram. In diagram (b) of fig. (\ref{flow}), $q_{-}^{'}=-q_{-}$.
Now the minus momenta of lines 2,3 and 4 can take positive or negative values, 
so more flow diagrams must be drawn to show these possibilities. Here, we make 
an approximation by setting $p_{1-}$ and $p_{1-}{'}$ to 0. 
This approximation is justified only in the
calculation of leading logarithms. 
Then the conservation of minus momenta applies that the direction of arrows on lines
1,2 and 3 must not be changed. 
Finally we have four flow diagrams for a box diagram. Diagram (d) is not 
allowed because at lower left vertex all the momenta are going in. 
In diagram (a), all the four arrows are in same direction so in each propagator
the loop momenta $q>0$. This implies that all the four poles belonging to each 
porpagator lies in lower half plane and if we enclose our contour by upper half 
plane contribution from this diagram is zero, same reason holds for 
diagram (b). For fig (c), the situation is different. Arrows on lines 1,2 and 3 
are in the same direction but the arrow on line 4 is in opposite direction to other 
three lines. Hence, only (c) contributes to the scattering amplitude corresponding to
the box diagram.

In diagram (c), the poles of line 1,2 and 3 lies in lower half plane but the pole 
of line 4 with propagator ${P_2-q}^2+i\epsilon$ lies in the upper half plane as we 
can see from sign of q. For flow diagram (a), either 
we can enclose the lower three poles or the upper one pole and it is easier 
to enclose one pole. Now, the scattering amplitude can be easily calculated. We
label the momenta as $p_1=\left(\omega,0,0,\omega\right)$ and 
$p_1=\left(\omega,0,0,-\omega\right)$ in the centre of mass frame, so that
the light cone momenta $p_{1-}=p_{2+}=0$, and denoting $q_-=2\omega x$, 
where the values of $x$ are specified
to lie between $0$ and $1$ from the conditions of positivity of line momenta along
the arrows of diagram (c). For the line of momentum $(-p_1'+q)^2 - m^2$, the denominator 
can be easily seen to be $sx$. Now, integration over the variable $x$ (which replaces
the integration over $q_-$ will produce the desired factor of ${\mbox{log}}s$, if we
cut off the $q_-$ integration to the order $\frac{1}{s}$. The final answer is
\be 
{\mathcal M}_{box}=-\frac{g^4s{\mbox{ln}}s}{4\pi}\int\frac{d^4q}{{(2\pi)}^4}
\frac{1}{\left(\qperp\right)^2
\left(\qperp-\d\right)^2}
\ee
\newpage
\bfi
\begin{center}
\mbox{\epsfig{file=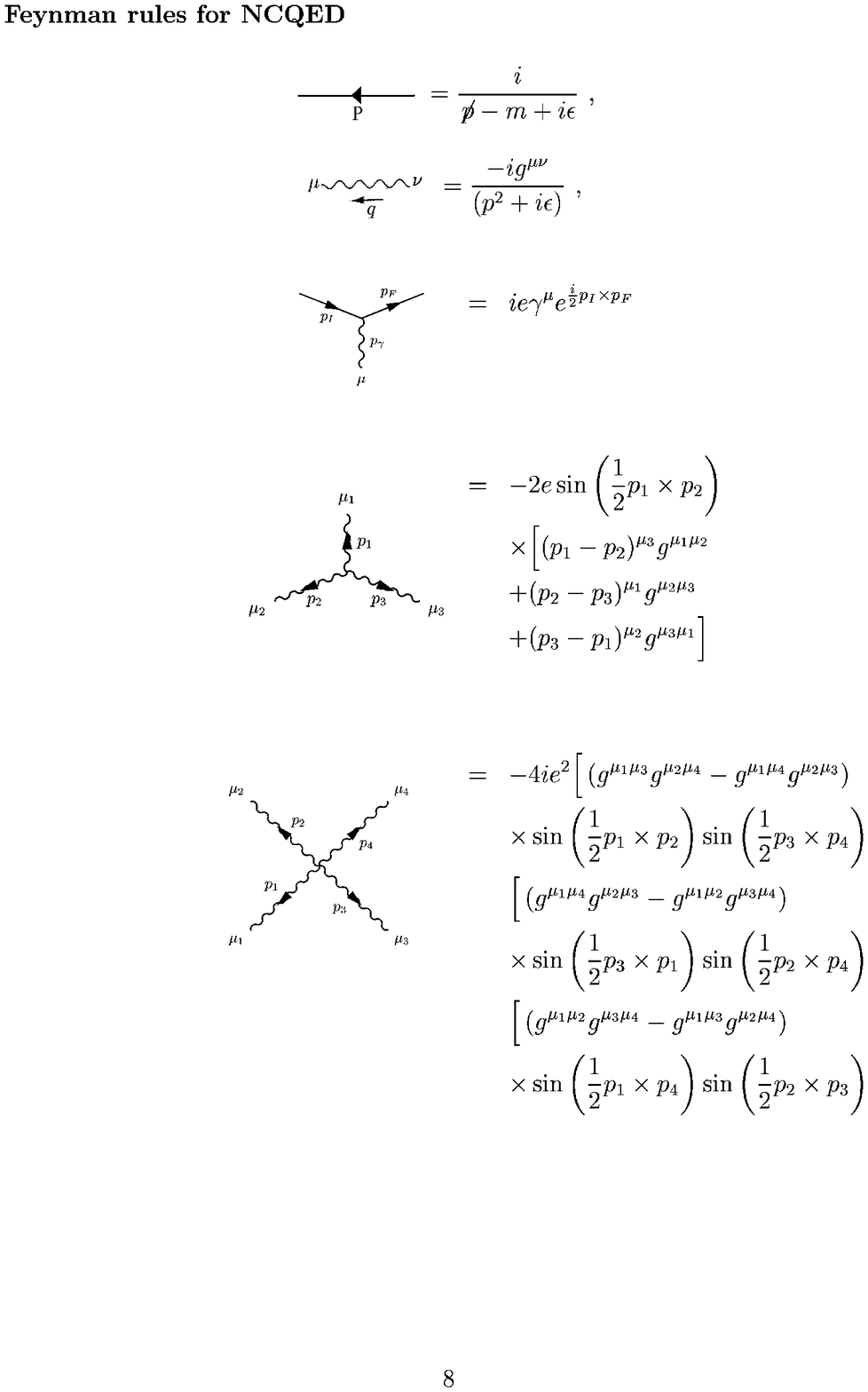,width=16.5truecm,angle=0}}
\label{feynman}
\end{center}
\efi

\bfi
\begin{center}
\mbox{\epsfig{file=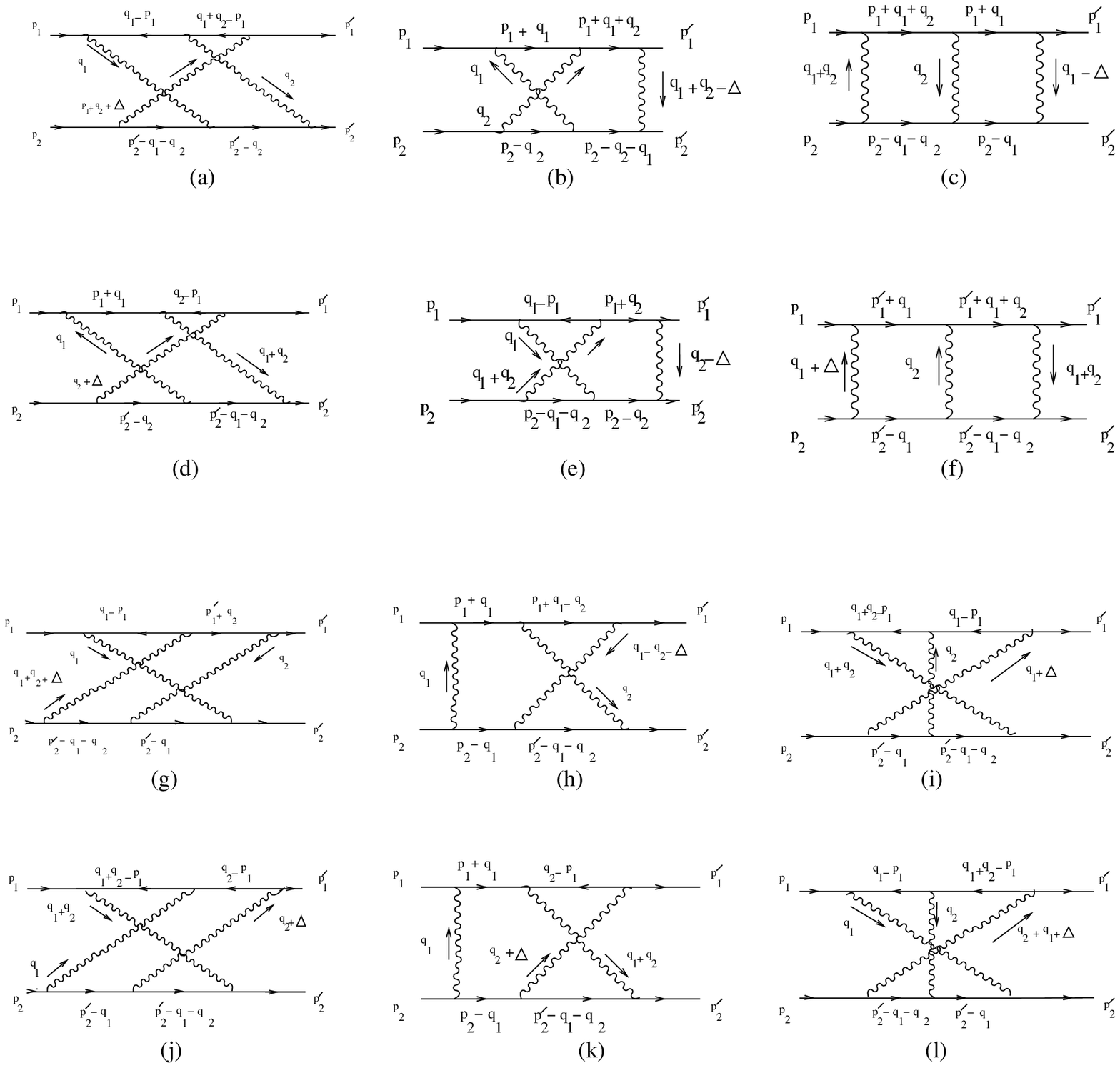,width=16.5truecm,angle=0}}
\caption{Flow diagrams contributing to the sixth order amplitude in e-e scattering.
Each of the diagrams in (\ref{ladder2}) has two flow diagrams, which have been
depicted in the figure with appropriate labelling.}
\label{flow3}
\end{center}
\efi

\end{document}